\documentclass[doublecol]{epl2} 

\usepackage{amsmath}
\usepackage{xr}
\usepackage{graphicx}

\makeatletter
\newcommand*{\addFileDependency}[1]{
  \typeout{(#1)}
  \@addtofilelist{#1}
  \IfFileExists{#1}{}{\typeout{No file #1.}}
}
\makeatother

\title{Engineering the sensitivity of macroscopic physical \\ systems to variations in the
fine-structure constant}
\shorttitle{Engineering the sensitivity of macroscopic physical systems to variations in the
fine-structure constant} 

\author{B. Zjawin\inst{1,2,*} \and M. Bober\inst{1} \and R. Ciury\l{}o\inst{1} \and D. Lisak\inst{1} \and M. Zawada\inst{1} \and P. Wcis\l{}o\inst{1,\dagger}}
\shortauthor{B. Zjawin \etal}

\institute{                    
  \inst{1} Institute of Physics, Faculty of Physics, Astronomy and Informatics, Nicolaus Copernicus University, Grudzi\c{a}dzka 5, 87-100 Toru\'n, Poland\\
  \inst{2} International Centre for Theory of Quantum Technologies, University of Gda\'nsk, 80-308 Gda\'nsk, Poland \\
  \inst{*} beata.zjawin@phdstud.ug.edu.pl\\
  \inst{\dagger} piotr.wcislo@umk.pl
}

\abstract{
Experiments aimed at searching for variations in the fine-structure constant $\alpha$ are based on spectroscopy of transitions in microscopic bound systems, such as atoms and ions, or resonances in optical cavities. The sensitivities of these systems to variations in $\alpha$ are typically on the order of unity and are fixed for a given system. For heavy atoms, highly charged ions and nuclear transitions, the sensitivity can be increased by benefiting from the relativistic effects and favorable arrangement of quantum states. This article proposes a new method for controlling the sensitivity factor of macroscopic physical systems. Specific concepts of optical cavities with tunable sensitivity to $\alpha$ are described. These systems show qualitatively different properties from those of previous studies of the sensitivity of macroscopic systems to variations in $\alpha$, in which the sensitivity was found to be fixed and fundamentally limited to an order of unity. Although possible experimental constraints attainable with the specific optical cavity arrangements proposed in this article do not yet exceed the present best constraints on $\alpha$ variations, this work paves the way for developing new approaches to searching for variations in the fundamental constants of physics.}

\begin{document}

\maketitle

\section{Introduction}
Although the standard model describes nature at the microscale with remarkable accuracy, it does not provide an explanation of the values of the parameters on which it is based, i.e., the fundamental constants. This has triggered a number of questions and hypotheses, including those about possible variations in the fundamental constants, which are also motivated by fundamental forces unification theories in which time and space variations in the fundamental constants naturally appear \cite{barrow2003constants}. 

Recently, considerable experimental effort (both laboratory experiments \cite{rosenband2008frequency,godun2014frequency,wcislo2016searching,wcislo2018first} and astrophysical observations \cite{bagdonaite2012stringent}) has been made to search for variability in the fundamental constants. In this paper, we focus on the fine-structure constant $\alpha$. We show that it is possible to design a macroscopic physical system for which the sensitivity to variations in $\alpha$ can be controlled. As a specific example, we theoretically consider a class of optical cavities with singular configurations for which the sensitivity of the frequency of the cavity modes to variations in $\alpha$ can be tuned, in principle, up to infinity. Although possible experimental constraints on $\alpha$ variations attainable with the specific designs proposed in this article would not exceed the present best constraints, our considerations show that the sensitivity factor does not need to be fixed for a given system and that macroscopic systems can have enhanced sensitivity to variations in $\alpha$. This introduces new perspectives on the search for variations in physical constants and the related tests of fundamental physics, such as dark matter searches \cite{derevianko2014hunting,arvanitaki2015searching,wcislo2016searching,wcislo2018first} and tests of extra-dimensional theories \cite{uzan2003fundamental}.

The detection limits of atomic, molecular, and optical experiments aimed at searching for $\alpha$ variations are determined by two factors: 1) the sensitivity of the considered frequency reference to variations in $\alpha$ and 2) the relative precision of the frequency measurements. The highest relative precision is achievable in the optical domain \cite{ushijima2015cryogenic,hansch}, e.g., with optical cavities approaching $10^{-17}$ in the $1$  to $100$ s time range \cite{matei20171,robinson2019crystalline,zhang2017ultrastable}. To quantify the sensitivity of a frequency reference to variations in $\alpha$, a dimensionless coefficient, $K_{\alpha}$, is defined as
  
\begin{equation}\label{K}
\frac{\delta\nu_0}{\nu_0}=K_{\alpha}\frac{\delta\alpha}{\alpha},
\end{equation}

\noindent
where $\nu_0$ is the frequency of the reference \cite{kozlov2018sensitivity,flambaum2009search}. The value of $K_\alpha$ depends on the choice of units; in this work, we use SI units unless otherwise stated. Searches for variations in $\alpha$ usually benefit from a comparison between different ultranarrow atomic or ionic optical transitions \cite{huntemann2014improved,hees2016searching,van2015search}. In general, the effective sensitivity coefficient for a comparison of two optical atomic transitions is on the order of unity or smaller and increases with the difference in the nuclear charges of the two atomic species due to relativistic corrections \cite{dzuba1999space,dzuba1999calculations,flambaum2009search}. Recently, a transition in Yb was proposed as a new frequency standard, having the highest sensitivity coefficient among optical atomic clock transitions, $K_\alpha=-13$ ($\widetilde{K}_\alpha=-15$ in atomic units) \cite{safronova2018two}. $K_\alpha$ can be further improved by applying  highly charged ions for which $K_\alpha$ increases with the ionization potential, reaching $K_\alpha=1142$ ($\widetilde{K}_\alpha=1140$ in atomic units) in Cf$^{16+}$  \cite{berengut2012optical}. Currently, the strongest $\alpha$ dependence has been predicted for a nuclear transition in $^{229}$Th for which $K_\alpha\approx-9000$ \cite{flambaum2006enhancednuclear,fadeev2020sensitivity}. We note that for some systems with transitions between nearly degenerate states, the sensitivity coefficient is strongly enhanced \cite{flambaum2006enhancedmol,angstmann2006narrow,flambaum2007enhanced,leefer2013new}. However, since the level spacing lies in the radio/microwave frequency range, the measurements based on such transitions do not benefit from ultrahigh relative precision relevant for optical metrology. In addition to atomic, ionic and nuclear transitions, macroscopic systems, such as optical resonators, can be used to search for variations in $\alpha$ \cite{stadnik2015searching,stadnik2016enhanced,wcislo2018first,wcislo2016searching}. Existing optical resonators with solid-state spacers are limited to $K_\alpha = 1$, and there have been no proposals on how to significantly enhance their $\alpha$ dependence. Moreover, for all the systems mentioned above, the sensitivity of a frequency reference to variations in $\alpha$ is fixed without capability to tune it.

In this work, we study macroscopic physical systems. As an example of such systems, we focus on optical resonators. We show that their sensitivity to variations in $\alpha$ can be controlled and enormously enhanced by replacing ordinary solid-state spacers (whose length is solely determined by electrostatic interactions) with classical macroscopic systems whose equilibrium points are determined by a balance between not only electrostatic but also magnetostatic interactions. We identify singularities in the static solutions at which $K_{\alpha}$ diverges to infinity. 

\section{Optical cavity design}

The frequency of the $N_{cav}$~-~th longitudinal mode of an optical cavity in vacuum is given by
\begin{equation}\label{f_SI_new_new}
\nu_{cav}=\frac{N_{cav} c}{2R},
\end{equation}
where $c$ is the speed of light and $R$ is the distance between the mirrors. In the case of ordinary optical resonators, $R$ is simply equal to the length of the solid-state spacer. Within the Born-Oppenheimer approximation, which includes only electrostatic interactions, the linear dimensions of any physical object (starting from atoms and molecules up to crystals and amorphous solids such as ultra-low expansion glass) scale as $\propto \alpha^{-1}$; see the Methods in \cite{wcislo2016searching} for a derivation in SI units and \cite{stadnik2016enhanced,lammerzahl2006search} for a derivation in a.u. Therefore, the length of the cavity spacer can be expressed as
\begin{equation}
    \label{RvsAlpha}
    R=\rho \alpha^{-1},    
\end{equation}
where $\rho$ is a factor that does not depend on $\alpha$. It follows from Equation~(\ref{f_SI_new_new}) and (\ref{RvsAlpha}) that 
\begin{equation}\label{f_SI_ordinary}
\nu_{cav}=\frac{N_{cav} c}{2\rho}\alpha.
\end{equation}
After substituting Equation~(\ref{f_SI_ordinary}) into Equation~(\ref{K}), one can simply show that $K_\alpha=1$ (in atomic units $\widetilde{K}_\alpha=-1$). This analysis was carried out in reference~\cite{wcislo2016searching} in SI units and in references~\cite{stadnik2015searching,stadnik2016enhanced} in atomic units. Considering the relativistic effects in solid-state structures leads to a deviation from $K_\alpha = 1$ on the order of $0.1$ for gold and lead, with less deviation for lighter elements such as those of cavity spacer materials \cite{stadnik2015searching,stadnik2016enhanced,pavsteka2018material}. This implies that $K_{\alpha}\approx 1$  for ordinary cavities.

\begin{figure}
\centering
\includegraphics[width=0.8\linewidth]{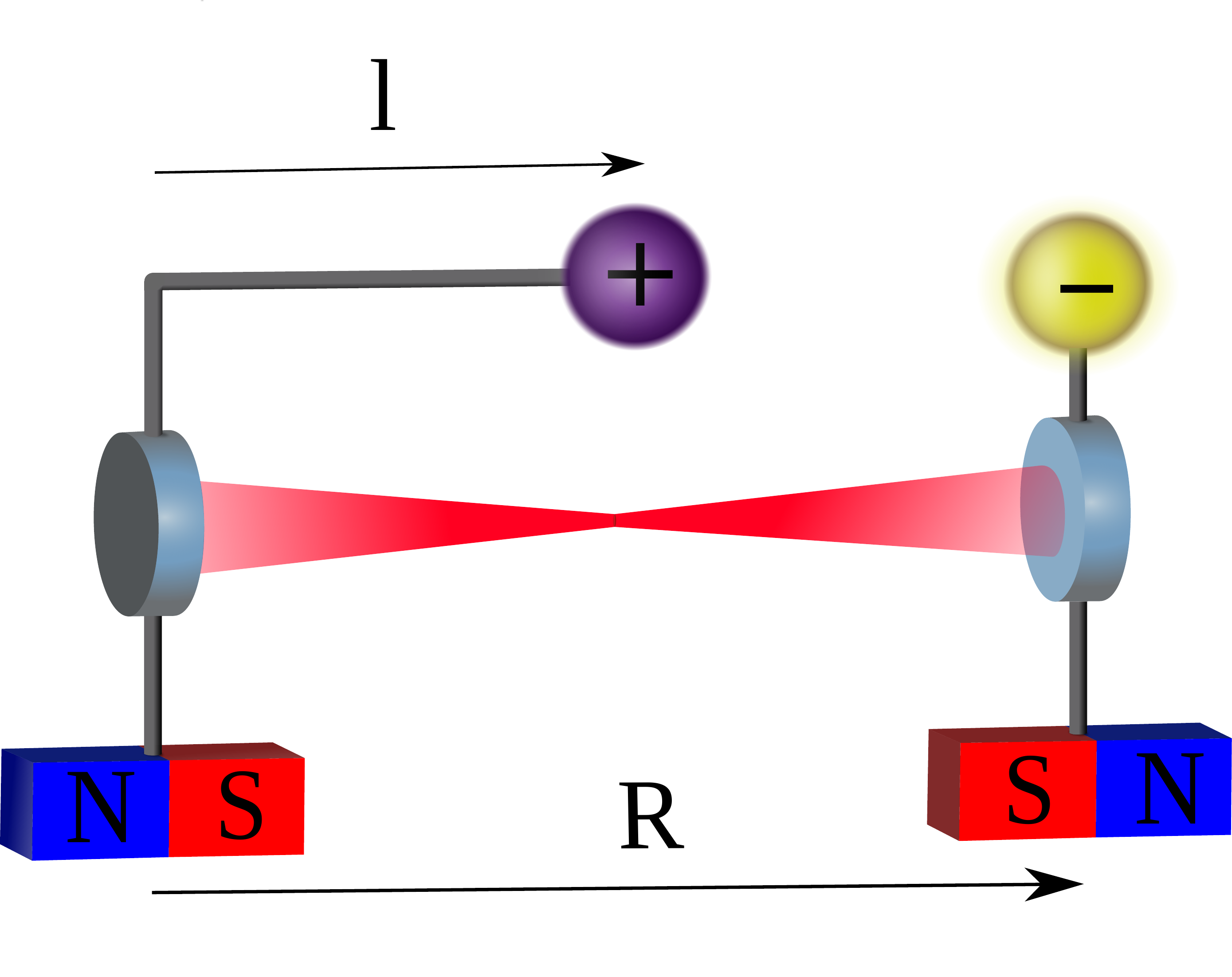}
\caption{The concept of an optical resonator with tunable sensitivity to fine-structure constant variations. The separation between the mirrors, $R$, is determined by the equilibrium between the electrostatic and magnetostatic interactions. The separation between the magnets is the same as the cavity length, $R$, and the separation between the charged spheres is $R-l$. The position of the equilibrium point is dictated by the $l/R_0$ ratio, where $l$ is the length of the offset rod and $R_0$ is the zero-offset equilibrium cavity length (the separation between the mirrors when no offset is applied and the forces are in balance). The mirrors can move only along the cavity axis; for simplicity, we do not present a mechanical system that limits movement in other dimensions. }\label{Fig1}
\end{figure}

We show how to engineer the sensitivity of optical resonators to variations in $\alpha$. Our method is based on replacing an ordinary solid-state spacer with a different mechanism that determines the distance between the mirrors, $R$. For an ordinary solid-state spacer, the $\alpha$-dependence of $R$ is given by Equation~(\ref{RvsAlpha}). Here, we show that the $\alpha$-dependence of $R$ can be arbitrarily tuned over a wide range when $R$ is determined not by the length of the solid-state spacer but by the balance between electrostatic and magnetostatic forces, as illustrated in Figure~\ref{Fig1}. Note that the relation in Equation~(\ref{f_SI_new_new}) is valid for any type of mechanism that sets the spacing of the mirrors, $R$, and once the $R(\alpha)$ dependence is calculated, the sensitivity coefficient $K_\alpha$ can be directly determined from Equation~(\ref{K}) and (\ref{f_SI_new_new}).

\subsection{The $\alpha$-dependencies of the electrostatic and magnetostatic forces}

To calculate the $\alpha$-dependence of $R$ for the arrangement shown in Figure~\ref{Fig1}, we start by discussing the $\alpha$~-~dependencies of the electrostatic force between two charged spheres, $F_Q$, and the magnetostatic force between two magnets, $F_M$. The electrostatic force can be expressed as
\begin{equation}\label{FQ_SI}
F_Q(r)=\frac{N_e^2e^2}{4\pi\varepsilon_0}r^{-2} =\xi_{Q}\alpha r^{-2},
\end{equation}
where $N_e$ is the number of elementary charges stored in each of the spheres, $e$ is the elementary charge, $\varepsilon_0$ is the vacuum permittivity, $\xi_{Q}=N_e^2\hbar c$, $\hbar$ is the reduced Planck's constant, $c$ is the speed of light and the fine-structure constant is given by 
\begin{equation}\label{eq:alpha_main}
\alpha~=~\frac{e^2}{4\pi\varepsilon_0\hbar c}.
\end{equation}
The magnetostatic force acting between two parallel and coaxial magnets, each having a magnetic moment $\mu$, can be written as

\begin{equation}\label{FM_SI}
F_M(r)=\frac{3\mu_0}{2\pi}\frac{\mu^2}{r^4}=\frac{3}{2\pi}\frac{1}{\varepsilon_0 c^2}\frac{\mu^2}{r^4}.
\end{equation}

\noindent
To analyze the dependence of a magnetic moment, $\mu$, on $\alpha$, we consider the example of a magnetic interaction originating from the intrinsic magnetic dipole moment of an electron, $\mu_s$, which can be written as

\begin{equation}\label{US}
\mu_s=-g_s \mu_B S,
\end{equation}

\noindent
where $g_s$, $\mu_B$ and $S$ are the electron $g$-factor, Bohr magneton and spin ($S=1/2$), respectively. The Bohr magneton is given by $\mu_B=(e\hbar)/(2m_e)$, where $m_e$ is the electron mass. Perturbation calculations show that $g_s$ can be expressed as

\begin{equation}\label{GS}
g_s=2+2 C_2\left(\frac{\alpha}{\pi}\right)+2 C_4\left(\frac{\alpha}{\pi}\right)^2+2 C_6\left(\frac{\alpha}{\pi}\right)^3+\cdots ,
\end{equation}

\noindent
where the Schwinger correction, $C_2$, equals $1/2$. The coefficients $C_i$, for $i=4,6$, are calculated analytically, and the coefficients $C_i$, for $i=8,10$, are evaluated numerically \cite{aoyama2012tenth}. Equation (\ref{GS}) shows that, to the first order, $g_s$ does not depend on $\alpha$, and the first leading term changes the effective sensitivity of $g_s$ to $\alpha$, $(d g_s / d \alpha) / (g_s / \alpha)$, by only $0.0012$. We ignore this small correction, which allows us to write $\mu_s=-2\frac{1}{2}\mu_B=-\mu_B$ and, hence, $\mu^2=N_{\mu}^2 \mu_B^2=N_{\mu}^2e^2\hbar^2/(2m_e)^2$ for $N_{\mu}$ contributing electrons. Therefore, the $\alpha$ dependence of the magnetostatic force in SI units is
\begin{equation}
F_M (r)=\frac{3}{2\pi c^2\epsilon_0}\frac{N_\mu^2e^2\hbar^2/(2m_e)^2}{r^4}=\frac{3\hbar^3}{2m_e^2c}\frac{N_\mu^2}{r^4}\left(\frac{e^2}{4\pi \epsilon_0\hbar c}\right).
\end{equation}
After substituting $\alpha~=~e^2/(4\pi\varepsilon_0\hbar c)$, we obtain the $\alpha$-dependence of $F_M$ in SI units
\begin{equation}
F_M(r)=\frac{3\hbar^3}{2m_e^2c}\frac{N_\mu^2}{r^4}\alpha=\xi_M \alpha r^{-4},
\label{FM_SI1}
\end{equation}
\noindent
where $\xi_{M}=(3\hbar^3N_\mu^2)/(2m_e^2c)$. The analysis provided here can be repeated for magnetic interactions originating from the orbital angular momentum, $\mu_l$, which leads to the same $\alpha$ dependence of the magnetostatic force. Thus, both forces electrostatic ($F_Q$) and magnetostatic ($F_M$) forces are proportional to $\alpha$ in SI units and depend on $r$ as $\propto r^{-2}$ and $\propto r^{-4}$, respectively.

\subsection{The equilibrium point of the cavity}

In the arrangement illustrated in Figure~\ref{Fig1} the separations between the magnets and between the mirrors are the same and are denoted as $R$. One of the charged spheres is placed on a solid-state rod that acts as an offset from the position of the magnet. Therefore, the separation between the charged spheres is $R-l$. The offset rod plays a central role in this configuration since it changes the local power-law $R$ dependence of the electrostatic contribution to the force acting on the mirrors, $F_{Q}=\xi_Q\alpha(R-l)^{-2}$. This force can be locally represented as $F_{Q} \propto R^{-N}$. Without the offset rod (for $l=0$), $N=2$. When a nonzero offset is applied ($l\neq0$), the power 

\begin{equation}\label{N}
N=2\frac{R}{R-l}
\end{equation}

\noindent
can be arbitrarily tuned in the range of 2 to infinity by changing the values of $R$ and $l$ (we consider here only the relevant $R>l$ case); see Appendix \ref{appendix:N} for the derivation. For the magnetostatic force, there is no offset rod; therefore, the force depends on $R$ as $F_M\propto R^{-4}$ ($N=4$). The two forces, $F_{M}$ and $F_{Q}$, are shown as a function of $R/R_{0}$ (black and green lines, respectively) in the upper panel in Figure~\ref{Fig2}. Since the result is shown on a log-log plot, the power $N$ in Equation~(\ref{N}) gives the local slope of the curves. The zero-offset equilibrium length, $R_{0}=\sqrt{\xi_M/\xi_Q}$, is the distance between the magnets when the electrostatic and magnetostatic forces are in balance with no applied offset ($l=0$). The value of the $l/R_0$ ratio determines the relative  position of the black and green curves in Figure~\ref{Fig2}, which determines the number of equilibrium points, i.e., the number of static solutions in which the forces are balanced. The balance is achieved when $F_Q(r=R-l)=F_M(r=R)$, that is, after substituting the expressions from Equation~(\ref{FQ_SI}) and (\ref{FM_SI1})
\begin{equation}
    \label{FQFMbalance}
    \xi_Q\frac{1}{(R-l)^2}=\xi_M\frac{1}{R^4}.
\end{equation}
After ignoring the solutions for $l>R$, which are irrelevant for this analysis, Equation~(\ref{FQFMbalance}) can be written in the form of a quadratic function: 
\begin{equation}
    \label{FbalanceQuadratic}
    R^2-R_0R+R_0l=0.
\end{equation}
Balance between the two forces is achieved for $R$ that are solutions of Equation~(\ref{FbalanceQuadratic}):
\begin{equation}\label{R_new}
R=\frac{1}{2}R_0 \left(1\mp\sqrt{1-4\frac{l}{R_0}}\right).\end{equation}
In the particular case of $l/R_0=0.25$, there is exactly one equilibrium point (the crossing point of the green and black solid lines in Figure~\ref{Fig2}). Smaller and larger values of $l/R_0$ lead to two (green dotted line in Figure~\ref{Fig2}) and zero (green dashed line in Figure~\ref{Fig2}) static solutions, respectively. When the number of static solutions is two, their stability depends on which force is chosen to be attractive. In the configuration presented in Figure~\ref{Fig1}, when the electrostatic force is attractive and the magnetostatic force is repulsive, the solution marked with the blue dot in the upper panel in Figure~\ref{Fig2} is stable, and the red-marked solution is unstable. For opposite signs of the forces, the situation is reversed: the red-marked solution is stable, and the blue-marked solution is unstable.

\section{Susceptibility to variations in $\alpha$}

The $\alpha$-dependence of $R$ can be directly determined from Equation~(\ref{R_new}). Since $R_0$ does not depend on $\alpha$, the $\alpha$-dependence arises only from the length of a solid-state spacer, $l$. As noted above (see the discussion before Equation~(\ref{RvsAlpha})), the linear dimensions of solid-state spacers are proportional to $\alpha^{-1}$; hence, $l$ can be written as 
\begin{equation}
    \label{eq:l}
    l=\lambda\alpha^{-1},
\end{equation}
where $\lambda$ is a parameter that does not depend on $\alpha$. It follows from Equation~(\ref{R_new}) and (\ref{eq:l}) that the $\alpha$-dependence of R is given by
\begin{equation}\label{Ralfa2}
R(\alpha)=\frac{1}{2}R_0 \left(1\mp\sqrt{1-4\frac{\lambda}{R_0}\alpha^{-1}}\right).
\end{equation}
Substituting this result into Equation~\eqref{f_SI_new_new} gives the $\alpha$ dependence of the frequency of a cavity mode:
\begin{equation}\label{nuAlfa2}
\nu_{cav}(\alpha)=\frac{N_{cav}c}{R_0} \frac{1}{1\mp\sqrt{1-4\frac{\lambda}{R_0}\alpha^{-1}}}.
\end{equation}
It follows from Equation~(\ref{K}) that the sensitivity coefficient can be evaluated as
\begin{equation}\label{Kformula}
K_\alpha=\frac{\alpha}{\nu_{cav}}\frac{d\nu_{cav}}{d\alpha}.
\end{equation}
Substituting the $\nu_{cav}(\alpha)$ function from Equation~(\ref{nuAlfa2}) into Equation~(\ref{Kformula}) gives (the details of this step are given in Appendix \ref{app:Kalpha})
\begin{equation}\label{K_a_SI_new}
K_{\alpha}=\frac{1}{2}\pm \frac{1}{2}\frac{1}{\sqrt{1-4\frac{l}{R_0}}}.
\end{equation}

\begin{figure}[ht]
\centering
\includegraphics[width=\linewidth]{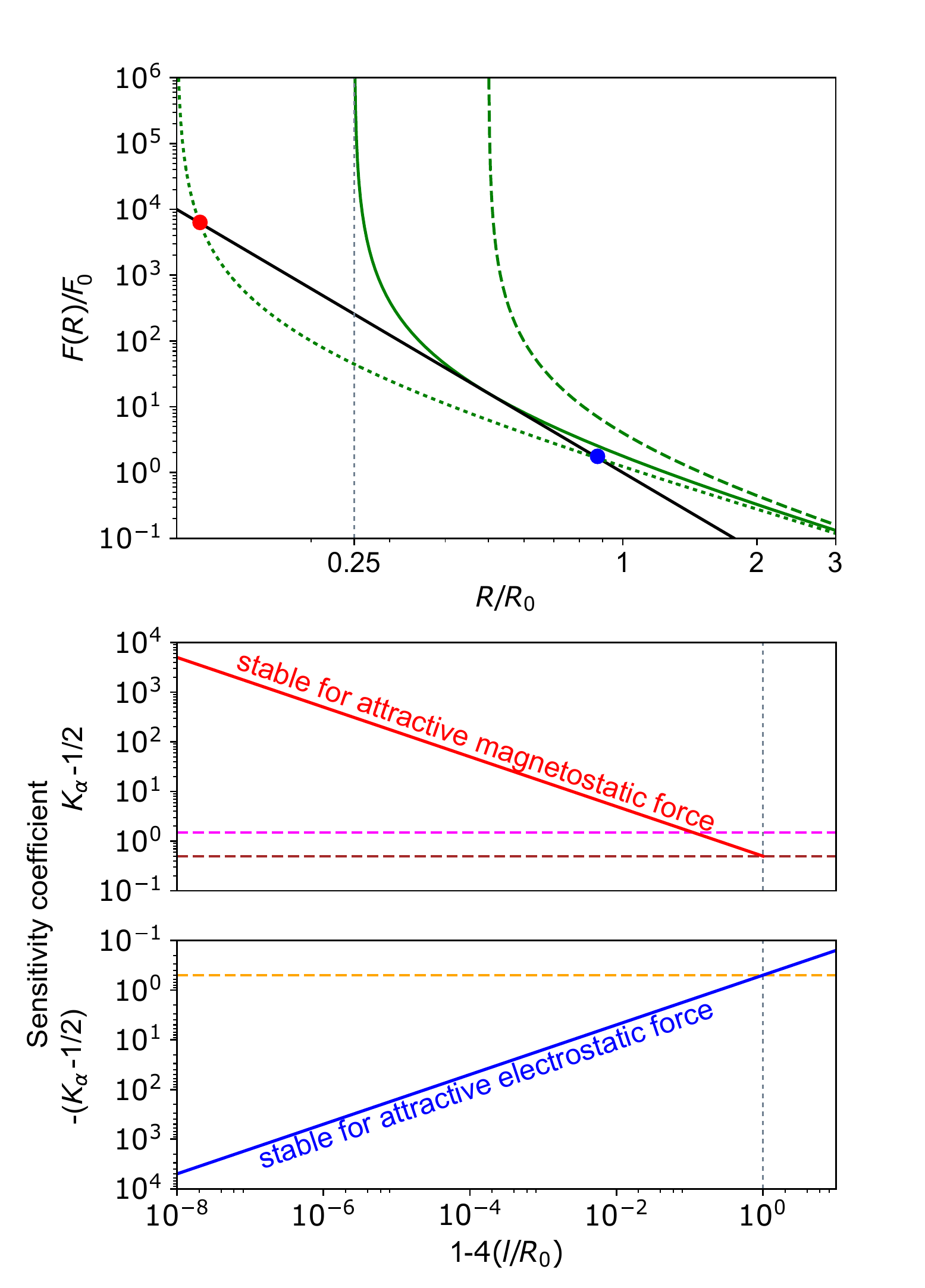}
\caption{Upper panel: electrostatic ($F_Q$) and magnetostatic ($F_M$) forces (green and black lines, respectively) as a function of the normalized cavity length, $R/R_0$. The forces are normalized to $F_0=F_M(R_0)=F_Q(R_0)$ for $l=0$. The offset-to-$R_0$ ratio, $l/R_{0}$, equals 0.1, 0.25 and 0.5 for the dotted, solid and dashed green lines, respectively. In the case of attractive electrostatic and repulsive magnetostatic forces, the blue and red dots correspond to the stable and unstable equilibrium points, respectively (the stability is exchanged for reversed signs of the forces). Bottom panel: the sensitivity coefficient as a function of the offset-rod length. The solution depicted with the blue line is stable for attractive electrostatic and repulsive magnetostatic forces. The solution shown with the red curve is stable for the reversed signs of the forces (the red and blue lines in the bottom panel correspond to the red and blue points in the upper panel). The orange line indicates the sensitivity for the case with no offset rod ($l=0$). The brown and magenta lines indicate the sensitivities for an ordinary cavity with a solid-state spacer and for an electronic transition in a nonrelativistic atom, respectively. $R_{0}$ is an equilibrium value of $R$ when there is no offset rod and $l=0$.}\label{Fig2}
\end{figure}

\noindent
A crucial point of our concept is that the value of $K_\alpha$ can be continuously tuned from $-\infty$ to $+\infty$ (excluding a small range from $K_\alpha=0.5$ to $1$) by changing the $l/R_0$ ratio. The adjustment of the $K_\alpha$ value towards the singular point (where $K_{\alpha}$ diverges to $\infty$) is illustrated in the bottom panel in Figure~\ref{Fig2}. The finite value of $K_{\alpha}$ corresponds to the stable configurations represented by the intersection of the dotted green line and the black line (blue or red point depending on the signs of the forces). When the $l/R_0$ ratio increases, the green dotted line moves towards the green solid line. The singular point occurs when the blue and red points converge to a single point, and the secant green dotted line becomes the tangent solid green line, that is, when $l/R_0=0.25$. In this configuration, $K_{\alpha}$ diverges to infinity. The $K_{\alpha}$ dependence on $l/R_0$ is presented in the bottom panel in Figure~\ref{Fig2}. The blue and red lines represent the stable and unstable branches, respectively, of the configuration presented in Figure~\ref{Fig1}, i.e., for the case of an attractive electrostatic force and a repulsive magnetostatic force. We also mark the values of $K_{\alpha}$ for some typical frequency references: the ordinary cavity with a solid-state spacer ($K_{\alpha}=1$, brown dashed line), the configuration without an offset ($K_{\alpha}=0$, orange dashed line) and a nonrelativistic atomic transition ($K_{\alpha}=2$, magenta dashed line). 

We showed that adding a macroscopic arrangement that combines electrostatic and magnetostatic interactions to a solid-state spacer results in strong enhancement of the sensitivity of an optical cavity to variations in $\alpha$. However, the configuration illustrated in Figure~\ref{Fig1} is not the only configuration that exhibits this feature. We discuss this finding in detail in Appendix \ref{sec:gravity}, where we demonstrate that the strong enhancement in the sensitivity to variations in $\alpha$ can be achieved in an arrangement that is based on a balance between electrostatic and gravitational interactions. Although a configuration that involves gravitational force is unrealistic from the point of view of experimental realization, it is noteworthy to mention it since it shows that the arrangement in Figure~\ref{Fig1} is not the only macroscopic physical system in which $K_{\alpha}$ is not fixed and can be arbitrarily tuned.

\section{Susceptibility to variations in other parameters}

In Section~3, we consider the sensitivity of the macroscopic system illustrated in Figure~\ref{Fig1} to variations in $\alpha$ under the assumption that other factors that may affect the cavity mode frequency do not change. Our analysis shows that macroscopic systems are not fundamentally limited to have fixed sensitivity factors. Although  the particular configurations proposed in this article would not exceed the present best experimental constraints on $\alpha$ variations, it is interesting to analyze their enhancement in sensitivity to potential sources of noise. We consider two examples: thermal noise of the mirror coatings and substrates and fluctuations in the offset-rod length.

\subsection{Example 1: thermal fluctuations of the mirror substrates and coatings}

For the configuration illustrated in Figure \ref{Fig1}, the equilibrium position $R$ is given by Equation \eqref{R_new}. If the thermal fluctuations of the mirror substrates and coatings are to be accounted for then the equilibrium point $R$ cannot be identified with the mirror spacing. In the denominator of Equation \eqref{f_SI_new_new}, $R$ should be replaced with $R+\epsilon$, where $\epsilon$ is an offset due to the way the mirrors are mounted, the mirror substrates thickness, etc. For simplicity, we may assume that mirror mounting is designed in a way that $\epsilon=0$, but we cannot assume that its noise, $\delta\epsilon$, is zero. Noise in $R$, $\delta R$, and $\delta\epsilon$ enter Equation \eqref{f_SI_new_new} independently; hence, the singular behavior of $R$ does not enhance the noise of the cavity mode frequency originating from thermal fluctuations of the mirror substrates and coatings ($\delta\epsilon$). Therefore, the enhancement in $K_\alpha$ directly improves the ratio of the hypothetical $alpha$-variation signal to this source of noise (with respect to an ordinary cavity with a solid-state spacer). It should be noted that the performance of the present best optical resonators is limited by the thermal noise of the mirror coatings \cite{robinson2021thermal}.

\subsection{Example 2: fluctuations in the length of the offset rod}

In this Section, we analyze another example of an instability source that exhibits qualitatively different behavior than the example in Section~4.1, that is, fluctuations in the offset rod length ($l$ in Figure~\ref{Fig1}). To quantify it, we define a dimensionless coefficient, $K_l$, as
\begin{equation}\label{K_l}
\frac{\delta\nu_0}{\nu_0}=K_{l}\frac{\delta l}{l}.
\end{equation}
In Section~3, Equation~\eqref{nuAlfa2}, we showed that for the configuration illustrated in Figure~1, the frequency of a cavity mode can be written as
\begin{equation}\label{eq:nu-l}
\nu_{cav}(l)=\frac{N_{cav}c}{R_0} \frac{1}{1\mp\sqrt{1-4\frac{l}{R_0}}}.
\end{equation}
By substituting it into the formula for the sensitivity coefficient, 
\begin{equation}\label{Kformula-l}
K_l=\frac{l}{\nu_{cav}}\frac{d\nu_{cav}}{dl},
\end{equation}
we obtain
\begin{equation}\label{K_l_AU}
K_{l}=-\frac{1}{2}\mp \frac{1}{2}\frac{1}{\sqrt{1-4\frac{l}{R_0}}}.
\end{equation}
For the configuration presented in Figure~1, the sensitivity coefficients $K_l$ and $K_{\alpha}$ have a similar form. It means that this specific configuration has enhanced sensitivities to both instability in $l$ and variations in $\alpha$.

In Appendix \ref{sec:gravity}, we discuss a different configuration in which the enhanced $K_{\alpha}$ does not entail an enhancement in $K_l$. For the configuration illustrated in Figure \ref{Fig3}, the sensitivity coefficient to the fluctuations in the length of the offset rod is  
\begin{equation}\label{K_l_FG}
K_{l}=-1,
\end{equation}
see Appendix \ref{sec:gravity} for the derivation. A comparison of Equation~(\ref{K_l_AU}) and (\ref{K_l_FG}) shows that the choice of a specific macroscopic system may dramatically change the enhancement of the cavity to a given quantity. 
\\

\section{Summary and discussion}
We proposed a construction of a macroscopic physical system with enhanced sensitivity to variations in $\alpha$. In contrast to various methods previously proposed for searches of variations in $\alpha$, the approach considered here allows the $\alpha$-sensitivity to be arbitrarily tuned, in principle, up to infinity. Instead of benefiting from favorable arrangements of quantum states in natural microscopic bound systems (such as atoms, molecules or ions), we show that $K_{\alpha}$ can be engineered by designing a synthetic macroscopic system that reveals some singularities in the equilibrium points. In particular, we calculated the singular points in which $K_{\alpha}$ diverges to infinity for the optical resonator configuration that includes electrostatic and magnetostatic interactions. Our result is qualitatively different from those of previous studies of the sensitivity of optical resonators to variations in $\alpha$, in which $K_{\alpha}$ was found to be fundamentally limited to an order of unity. This approach opens a way for further searches with other arrangements that would be possible from the perspective of an experimental realization. Finally, the concept demonstrated here has the potential to be extended to other fundamental constants, e.g., the proton-to-electron mass ratio.

\medskip

\medskip
\textbf{Acknowledgements} \par 
BZ is supported by the Foundation for Polish Science (IRAP project, ICTQT, contract no. 2018/MAB/5, co-financed by EU within Smart Growth Operational Programme). The ’International Centre for Theory of Quantum Technologies’ project (contract no. MAB/2018/5) is carried out within the International Research Agendas Programme of the Foundation for Polish Science co-financed by the European Union from the funds of the Smart Growth Operational Programme, axis IV: Increasing the research potential (Measure 4.3).
PW is supported by the National Science Centre, Poland, project no. 2019/35/B/ST2/01118. MB is supported by the National Science Centre, Poland, project no. 2015/19/D/ST2/02195. DL is supported by the Polish National Science Centre project no. 2015/18/E/ST2/00585. 
The contribution of MZ is supported by the National Science Centre, Poland, under the QuantERA Programme, no. 2017/25/Z/ST2/03021. 
The project "A next-generation worldwide quantum sensor network with optical atomic clocks" is carried out within the TEAM IV Programme of the Foundation for Polish Science cofinanced by the European Union under the European Regional Development Fund.
Support has been received from the project EMPIR 17FUN03 USOQS. This project has received
funding from the EMPIR Programme cofinanced by the Participating States and from the
European Union’s Horizon 2020 Research and Innovation Programme.
The research is a part of the program of the National Laboratory FAMO (KL FAMO) in Toru\'n, Poland, and is supported by a subsidy from the Polish Ministry of Science and Higher Education.
\newline
\newline

This is the Accepted Manuscript version of an article accepted for publication in Europhysics Letters. IOP Publishing Ltd is not responsible for any errors or omissions in this version of the manuscript or any version derived from it. This Accepted Manuscript is published under a CC BY licence. The Version of Record is available online at 10.1209/0295-5075/ac3da3.

\bibliography{bibliography}
\bibliographystyle{unsrt.bst}

\newpage

\medskip
\textbf{Appendix} \par 
\section{Power-law approximation of the electrostatic force}\label{appendix:N}

The electrostatic force acting on the mirrors in the configuration illustrated in Figure~\ref{Fig1} is given by $F_{Q}=\xi_Q\alpha(R-~l)^{-2}$. To see how the offset-rod length $l$ changes the local power-law $R$ dependence of the force, $F_Q$ can be locally represented as $F_{Q} \propto R^{-N}$. Without the offset rod ($l=0$), the force can be written as $F_{Q}=\xi_Q\alpha R^{-2}$, and we recover $N=2$. When a nonzero offset is applied ($l\neq0$), the force can be locally approximated as $\xi_Q\alpha(R-l)^{-2} \propto R^{-N}$, where the power $N$ can be extracted from the following equation:
\begin{equation}
  \frac{dF_Q}{F}=-N\frac{dR}{R}.
\end{equation}
Therefore, $N$ is given by
\begin{equation}
  N=-\frac{R}{F_Q} \frac{dF_Q}{dR},
\end{equation}
which gives the result presented in Equation~\eqref{N}:
\begin{equation}
N=2\frac{R}{R-l}.
\end{equation}

\section{The sensitivity coefficient $K_\alpha$}
\label{app:Kalpha}

In the main text, we provided a derivation (Equation~(\ref{FQFMbalance})-(\ref{K_a_SI_new})) of the formula for the sensitivity coefficient $K_\alpha$ (Equation~(\ref{K_a_SI_new})) for the arrangement in Figure~\ref{Fig1}. In this Supporting Information, we explain in more detail the step from Equation~(\ref{nuAlfa2}) and (\ref{Kformula}) to Equation~(\ref{K_a_SI_new}). Evaluation of the derivative from Equation~(\ref{Kformula}) is straightforward 
\begin{equation}
\frac{d}{d\alpha}\nu_{cav}(\alpha)=\pm\frac{N_{cav}c}{R_0}\frac{\frac{4\lambda}{R_0 \alpha}}{2\alpha\sqrt{1-\frac{4\lambda}{R_0 \alpha}}\left(1\mp\sqrt{1-\frac{4\lambda}{R_0 \alpha}}\right)^2}.
\end{equation}
To simplify the formula from Equation~(\ref{Kformula}) to the form of Equation~(\ref{K_a_SI_new}), a few additional steps are required:
\begin{equation}
K_\alpha=\pm\frac{1}{2}\frac{\frac{4\lambda}{R_0 \alpha}}{\sqrt{1-\frac{4\lambda}{R_0 \alpha}}\left(1\mp\sqrt{1-\frac{4\lambda}{R_0 \alpha}}\right)}    
\end{equation}
\begin{equation}
=\mp\frac{1}{2}\frac{\left(1-\frac{4\lambda}{R_0 \alpha}\right)-1}{\sqrt{1-\frac{4\lambda}{R_0 \alpha}}\left(1\mp\sqrt{1-\frac{4\lambda}{R_0 \alpha}}\right)}    
\end{equation}
\begin{equation}
=\frac{1}{2}\frac{\left(\sqrt{1-\frac{4\lambda}{R_0 \alpha}}\pm 1\right)\left(\sqrt{1-\frac{4\lambda}{R_0 \alpha}}\mp 1\right)}{\sqrt{1-\frac{4\lambda}{R_0 \alpha}}\left(\sqrt{1-\frac{4\lambda}{R_0 \alpha}}\mp 1\right)}    
\end{equation}
\begin{equation}
=\frac{1}{2}\frac{\sqrt{1-\frac{4\lambda}{R_0 \alpha}}\pm 1}{\sqrt{1-\frac{4\lambda}{R_0 \alpha}}}    
\end{equation}
\begin{equation}
=\frac{1}{2}\pm\frac{1}{2}\frac{1}{\sqrt{1-\frac{4\lambda}{R_0 \alpha}}}.    
\end{equation}

\section{Choice of units} 

Although $K_\alpha$ is a dimensionless coefficient, its value depends on the choice of units \cite{kozlov2018sensitivity-app}. In this work, we use SI units, but the analysis can be performed in any other system of units and will lead to exactly the same prediction of an actual experiment. In this section, we repeat the analysis from the main text but use atomic units (a.u.). We start by discussing the conversion factors between SI and a.u. for several physical quantities that are relevant for our analysis. The quantities in a.u. and SI units are denoted by a corresponding symbol with and without a tilde, respectively.

The fine-structure constant, $\alpha$, in SI units is expressed as 
\begin{equation}
    \alpha=\frac{e^2}{4\pi\varepsilon_0\hbar c},
    \label{eq:alpha}
\end{equation}
where $\varepsilon_0$ is the vacuum permittivity, $e$ is the elementary charge, $c$ is the speed of light and $\hbar$ is the reduced Planck's constant. The length unit in a.u. is bohr, which, in SI, units can be evaluated from the following expression:
\begin{equation}
     a_0=\frac{4\pi\varepsilon_0\hbar^2}{m_e e^2} =\left(\frac{\hbar}{m_e c}\right)\alpha^{-1},
     \label{eq:a0}
\end{equation}
where $m_e$ is the electron mass. Therefore, the relation between a length expressed in SI units, $r$, and in a.u., $\tilde{r}$, is
\begin{equation}
    r=\left(\frac{\hbar}{m_e c}\right)\alpha^{-1}\tilde{r}.
    \label{eq:r}
\end{equation}
The energy unit is hartree,
 \begin{equation}
     E_h=\frac{m_e e^4}{(4\pi\varepsilon_0\hbar)^2}=\left(m_ec^2\right)\alpha^2.
     \label{eq:Eh}
 \end{equation}
By combining Equation~(\ref{eq:a0}) and (\ref{eq:Eh}), we get the unit of force,
  \begin{equation}
\frac{E_h}{a_0}=\frac{m_e^2e^6}{(4 \pi\varepsilon_0)^3\hbar^4}=\left(\frac{m_e^2c^3}{\hbar}\right)\alpha^3.
  \end{equation}
Therefore, the relation between a force expressed in SI units, $F$, and in a.u., $\tilde{F}$, is  
\begin{equation}
      F=\left(\frac{m_e^2c^3}{\hbar}\right)\alpha^3\tilde{F}.
      \label{eq:F}
  \end{equation}

Equation~(\ref{eq:r}) and (\ref{eq:F}) are the rules needed to transform the physical quantities used in our work from SI units to a.u.

The electrostatic force between two charged spheres in SI units is
\begin{equation}
F_Q(r)=\frac{1}{4\pi\varepsilon_0}\frac{q^2}{r^2},
\end{equation}
where $q=e N_e$ is the charge stored at each of the spheres, $e$ is the elementary charge and $N_e$ is the number of elementary charges at each sphere. The explicit dependence of $F_Q$ on $\alpha$, which is given by Equation~\eqref{FQ_SI}, is obtained by using the $\alpha$ expression in SI units, given in Equation~(\ref{eq:alpha_main})
\begin{equation}\label{eq:FQ_SI_alpha}
F_Q (r)=\left(\hbar c\frac{N_e^2}{r^2}\right) \alpha.
\end{equation}
By directly substituting Equation~(\ref{eq:r}), (\ref{eq:F}) and (\ref{eq:FQ_SI_alpha}), we obtain
\begin{equation}
    \left(\frac{m_e^2c^3}{\hbar}\right)\alpha^3\tilde{F}_Q (\tilde{r})=\left(\hbar c\frac{N_e^2}{\left(\frac{\hbar}{m_e c}\right)^2\alpha^{-2}\tilde{r}^2}\right)\alpha.
\end{equation}
After simple rearranging and ordering the factors, we obtain the expression for the electrostatic force in atomic units
\begin{equation}\label{FQ_AU}
\tilde{F}_{Q} (\tilde{r})=\tilde{\xi}_Q\frac{1}{\tilde{r}^{2}},
\end{equation}
where $\tilde{\xi}_Q=N_e^2$.

The magnetostatic force between two parallel and coaxial magnets (each having a magnetic dipole moment $\mu$) in SI units was derived in the main text and can be written as (see Equation~\ref{FM_SI1})
\begin{equation}
F_M (r)=\frac{3\hbar^3}{2m_e^2c}\frac{N_\mu^2}{r^4}\alpha.
\label{FM-alpha}
\end{equation}
Following the approach adopted for the electrostatic force case, the conversion from SI units to a.u. is achieved by substituting Equation~(\ref{eq:r}) and (\ref{eq:F}) to obtain
\begin{equation}
    \left(\frac{m_e^2c^3}{\hbar}\right)\alpha^3\tilde{F}_M (\tilde{r})=\frac{3}{2} \frac{\hbar^3}{ m_e^2 c} \frac{N_\mu^2}{\left(\frac{\hbar}{m_e c}\right)^4\alpha^{-4}\tilde{r}^4}\alpha.
\end{equation}
After simple rearrangements, we obtain the $\alpha$-dependence of $F_M$ in a.u.
\begin{equation}\label{FM_AU1}
\tilde{F}_{M} (\tilde{r})=\left(\frac{3}{2}\frac{N_\mu^2}{\tilde{r}^{4}}\right)\alpha^2 =\tilde{\xi}_M\alpha^2\frac{1}{\tilde{r}^4},
\end{equation}
where $\tilde{\xi}_{M}=\frac{3N_\mu^2}{2}$.

The balance condition for the arrangement from Figure~\ref{Fig1} in a.u., $\tilde{F}_Q(\tilde{r}=\tilde{R}-\tilde{l})=\tilde{F}_M(\tilde{r}=\tilde{R})$, takes the following form:
\begin{equation}
    \tilde{\xi}_Q\frac{1}{(\tilde{R}-\tilde{l})^2}=\tilde{\xi}_M\alpha^2\frac{1}{\tilde{R}^4}.
\end{equation}
Assuming $\tilde{R}>\tilde{l}$, the balance condition can be written in the form of a quadratic function
\begin{equation}
    \label{eq:quadraticAU}
    \tilde{R}^2-\tilde{R}_0\tilde{R}+\tilde{R}_0\tilde{l}=0,
\end{equation}
where $\tilde{R}_0=n\alpha$, and $n=\sqrt{3/2}(N_\mu/N_e)$ is a dimensionless factor. The solution of Equation~(\ref{eq:quadraticAU}) is
\begin{equation}
    \tilde{R}=\frac{\tilde{R}_0}{2}\left(1\mp\sqrt{1-4\frac{\tilde{l}}{\tilde{R}_0}}\right).
\end{equation}
Hence, the $\alpha$-dependence of $\tilde{R}$ can be written as
\begin{equation}
    \label{RalphaDependence}
    \tilde{R}(\alpha)=\frac{n\alpha}{2}\left(1\mp\sqrt{1-4\frac{\tilde{l}}{n}\alpha^{-1}}\right).
\end{equation}
The unit of time in a.u. is $a_0/(\alpha c)$; hence, the frequencies of the cavity mode in SI units and a.u. are related by 
\begin{equation}
    \label{RtildeAlpha}
    \nu_{cav}=\frac{\alpha c}{a_0}\tilde{\nu}_{cav}.
\end{equation}
The length transforms as $R=a_0\tilde{R}$; hence, Equation~(\ref{f_SI_new_new}) can be expressed in a.u. as 
\begin{equation}\label{f_au}
\widetilde{\nu}_{cav}=\frac{N_{cav} }{2\widetilde{R}}\alpha^{-1}.
\end{equation}
The $\alpha$ dependence of $\widetilde{\nu}_{cav}$ can be calculated by merging Equation~(\ref{RalphaDependence}) and (\ref{f_au})
\begin{equation}
    \tilde{\nu}_{cav}(\alpha)=\frac{N_{cav}}{n}\alpha^{-2}\left(1\mp\sqrt{1-4\frac{\tilde{l}}{n}\alpha^{-1}}\right)^{-1}.
\end{equation}
Equation~(\ref{Kformula}) has the same form in SI units and a.u., which gives the sensitivity coefficient in a.u.

\begin{equation}\label{K_a_AU}
\widetilde{K}_{\alpha}=-\frac{3}{2}\pm \frac{1}{2}\frac{1}{\sqrt{1-4\frac{\widetilde{l}}{\widetilde{R}_0}}}.
\end{equation}
It clearly emerges from Equation~(\ref{K_a_AU}) and~(\ref{K_a_SI_new}) that the $K_{\alpha}$ values in atomic and  SI units are different. However, in an actual experiment, the physically meaningful quantity that can be measured is a dimensionless ratio of the frequency of the cavity considered here, $\nu_{cav}$, and the frequency of some other references, $\nu_{ref}$:

\begin{equation}\label{K_lab}
\frac{\delta(\nu_{cav}/\nu_{ref})}{(\nu_{cav}/\nu_{ref})}=(K_{\alpha}-K_{\alpha}^{ref})\frac{\delta\alpha}{\alpha},
\end{equation}
where $K_{\alpha}^{ref}$ is the sensitivity of the second reference to variations in $\alpha$. Equation~(\ref{K_lab}) has the same form in SI units and atomic units. The values of the individual sensitivities $K_{\alpha}$ and $K_{\alpha}^{ref}$ can differ in different choices of units; however, the effective sensitivity coefficient $K_{\alpha}-K_{\alpha}^{ref}$ is independent of the choice of units. As an example, we choose an ordinary cavity with a solid-state spacer as a second reference. For an ordinary cavity $K_{\alpha}^{ref}=1$ in SI units and $\widetilde{K}_{\alpha}^{ref}=-1$ a.u. \cite{kozlov2018sensitivity-app}. By substituting Equation~(\ref{K_a_AU}) and (\ref{K_a_SI_new}) into Equation~(\ref{K_lab}), we obtain the effective sensitivity coefficient:
 \begin{equation}\label{K_a_lab}
 K_{\alpha}-K_{\alpha}^{ref}= \widetilde{K}_{\alpha}-\widetilde{K}_{\alpha}^{ref}=\frac{1}{2}\mp \frac{1}{2}\frac{1}{\sqrt{1-4\frac{l}{R_0}}},
 \end{equation}
which takes the same value in both systems of units. Note that the offset-to-$R_0$ ratio has the same $\alpha$~-~dependence in SI units, $l/R_0 \propto \alpha^{-1}$, and in a.u., $\tilde{l}/\tilde{R_0} \propto \alpha^{-1}$. This example shows that if a comparison of two frequency references is considered, instead of a single frequency reference, the sensitivity coefficient is independent of the choice of units.

\section{Configuration with gravitational force}\label{sec:gravity}

\begin{figure}
\centering
\includegraphics[width=0.4\linewidth]{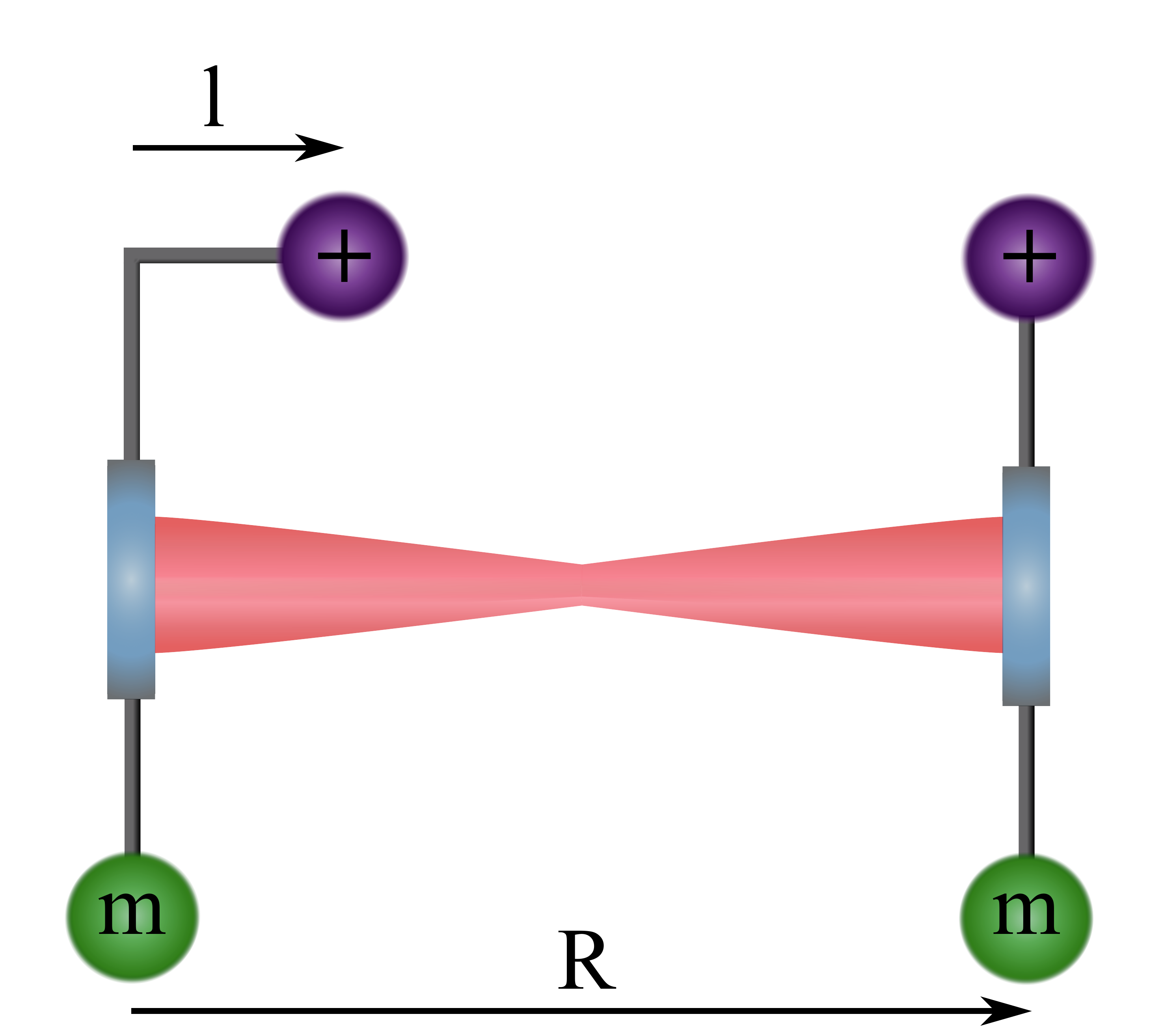}
\caption{Realization of an optical resonator with enhanced sensitivity to fine-structure constant variations. The length of the resonator, $R$, is determined by the equilibrium between the electrostatic and gravitational interactions. The separations between the masses and the mirrors, $R$, are the same, and the separation between the charged spheres is $R-l$. The position of the equilibrium point is set by $\sqrt{\alpha}\xi$, where $\xi$ reflects the charge stored at each of the spheres and the mass, $m$. For simplicity, a mechanical system that limits the mirror movement in dimensions other than the cavity axis is not presented. }\label{Fig3}
\end{figure}

The configuration illustrated in Figure~\ref{Fig1} is a particular case of enhancing the sensitivity of an optical resonator to $\alpha$ variations that involves electrostatic and magnetostatic interactions. One may imagine other configurations that include different forces (for example, gravitational or electrodynamic forces). In this Supporting Information, we discuss one other possibility that reveals some important qualitatively different features. We consider a model system that involves electrostatic and gravitational interactions; see Figure~\ref{Fig3}. Although such a configuration is unrealistic from the point of view of experimental realization at the moment (the gravitational force is far too weak to be in equilibrium with electrostatic force), it constitutes an example of a different configuration that also reveals enhanced $\alpha$ sensitivity and possesses qualitatively different features.

\begin{figure}[htp]
\centering
\includegraphics[width=0.7\linewidth]{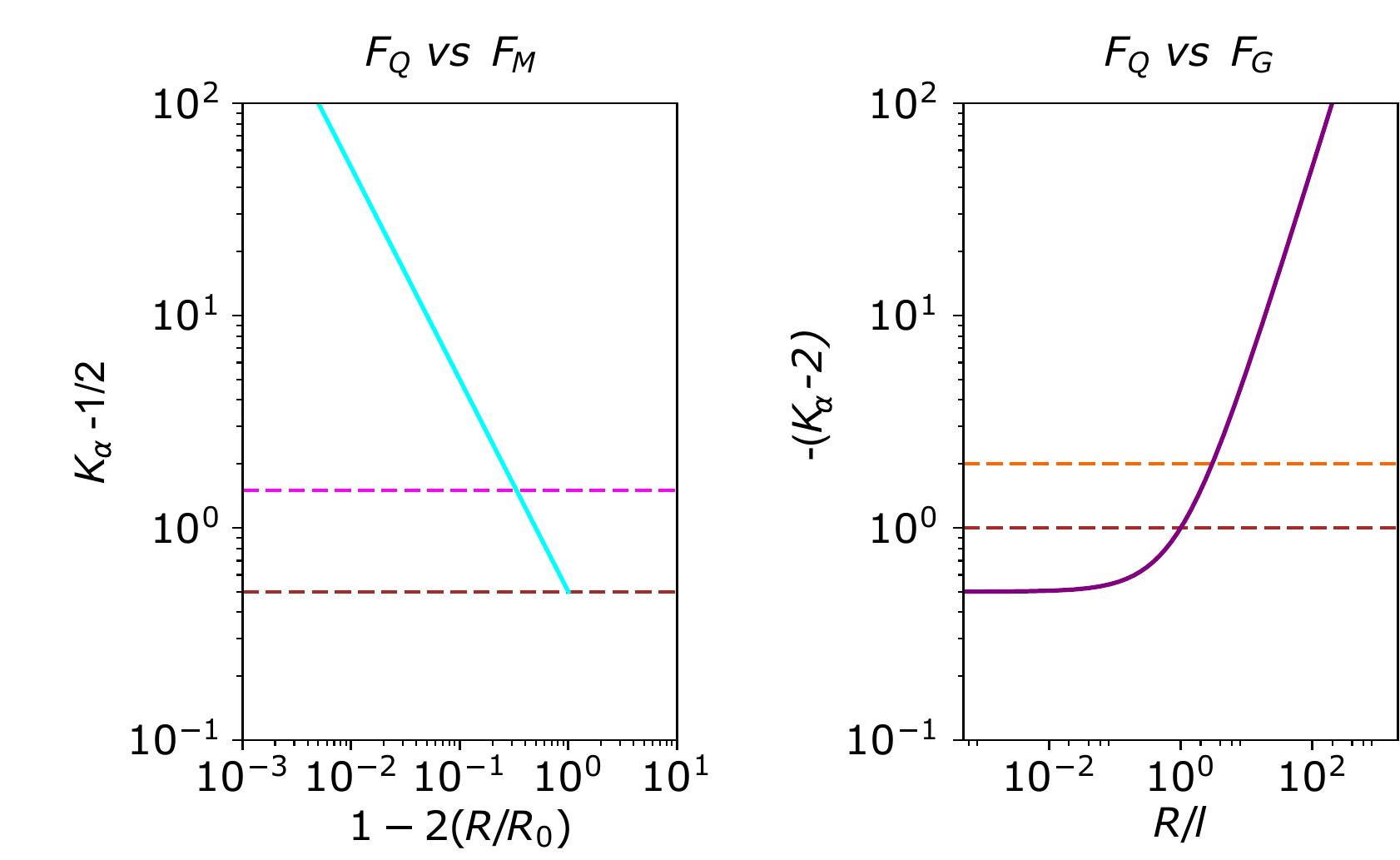}
\caption{Left panel: $K_{\alpha}$ as a function of the normalized cavity length, $R/R_0$ (cyan line), for the model system in which the length of the cavity is determined by the balance between $F_Q$ and $F_M$. Right panel: $K_{\alpha}$ as a function of the cavity length and the offset-rod length, $R/l$ (purple line), for the case employing $F_Q$ and $F_G$. The orange, brown and magenta lines indicate the sensitivity for the $F_Q$ vs $F_M$ case with no offset rod ($l=0$), an ordinary cavity with a solid-state spacer and an electronic transition in a nonrelativistic atom, respectively.}\label{Fig4}
\end{figure}

The gravitational force acting between two objects with masses $m$ is
\begin{equation}\label{FG_SI}
F_G (r)=G\frac{m^2}{r^2},
\end{equation}
where $G$ is the gravitational constant. We denote $\xi_G=~G m^2$. The gravitational force clearly does not depend on $\alpha$ in SI units. In the configuration illustrated in Figure~\ref{Fig3}, the separations between the mirrors and the attracting masses are equal and denoted by $R$. The separation between the charged spheres is $R-l$. Analogous to the configuration presented in Figure~\ref{Fig1}, the offset-rod length, $l$, is crucial since it changes the local power-law $R$ dependence of the electrostatic contribution to the force between the mirrors. The derivation of $K_{\alpha}$ that corresponds to this configuration is the same as the one performed in the main text for the configuration with magnetostatic force. Here, balance between the forces is achieved when $F_Q(r=R-l)=F_G(r=R)$, that is, when
\begin{equation}
    \label{Fbalance}
    \xi_Q \alpha \frac{1}{(R-l)^2}=\xi_G\frac{1}{R^2}.
\end{equation}
Balance is achieved for the following value of $R$:
\begin{equation}\label{R_FG}
R=\frac{1}{1-\sqrt{\alpha}\xi}l,
\end{equation}
where $\xi=\sqrt{\xi_Q/\xi_G}$. We consider here only the relevant $R > l$ case, which corresponds to $1>\sqrt{\alpha}\xi > 0$. Substituting the value of $R$ given by Equation~\eqref{R_FG} into Equation~\eqref{f_SI_new_new} gives the $\alpha$ dependence of the cavity modes:
\begin{equation}\label{nuAlfa3}
\nu_{cav}(\alpha)=\frac{N_{cav}c}{2 \lambda \alpha^{-1}} (1-\sqrt{\alpha}\xi),
\end{equation}
where we used the fact that the offset rod length scales inversely with $\alpha$, $l=\lambda \alpha^{-1}$. The sensitivity coefficient is then given by Eq~\eqref{K}. $K_{\alpha}$ for this configuration, in SI units, can be written as
\begin{equation}\label{K_a_QG_SI}
K_{\alpha}=\frac{1}{2\sqrt{\alpha}\xi-2}+\frac{3}{2}.
\end{equation}
In a similar manner to the configuration employing $F_Q$ and $F_M$, we identify a singular solution for $\sqrt{\alpha}\xi\rightarrow 1$ at which $K_{\alpha}$ diverges to infinity. There is, however, one important feature that distinguishes the two configurations presented in Figures~\ref{Fig1} and~\ref{Fig3}. The enhancement of the $K_{\alpha}$ coefficient for the $F_Q$ vs $F_G$ configuration occurs over a wide range of $R$ (i.e., $K_{\alpha}$ diverges to -$\infty$ for $R/l\rightarrow\infty$), while for the $F_Q$ vs $F_M$ configuration, it occurs in a narrow range of $R/R_0=0.5$; see Figure~\ref{Fig4}. 

We now examine the sensitivity of the configuration illustrated in Figure~\ref{Fig3} to variations in $l$. The cavity modes depend on $l$ as follows:
\begin{equation}\label{eq:nu-FG-l}
\nu_{cav}(l)=\frac{N_{cav}c}{2 l} (1-\sqrt{\alpha}\xi).
\end{equation}
It is straightforward to see that $K_l=-1$ in this case. As a consequence of $K_l$ being constant, the noise caused by $l$ instabilities will not be enhanced in the singular solutions where $K_{\alpha}$ diverges to infinity.

\end{document}